\documentclass[a4paper]{jpconf}
\usepackage{graphicx}

\begin{document}

\title{High-resolution Millimeter-VLBI Imaging of Sgr~A*}

\author{Zhi-Qiang Shen\footnote[1]{for collaborators see
acknowledgments}}

\address{Shanghai Astronomical Observatory, 80 Nandan Road, Shanghai 200030, China}
\ead{zshen@shao.ac.cn}

\begin{abstract}
We present the highest resolution VLBI imaging observations of
Sgr~A* made at both 7 and 3.5 mm. These data reveal
wavelength-dependent intrinsic sizes with an intrinsic emitting
region at 3.5 mm of about 1 AU (at a distance of 8 kpc to the
Galactic Center). When combined with the lower limit on the mass
of Sgr~A*, these size measurements provide strong evidence that
Sgr~A* is a super-massive black hole. We also detected a
structural variation which results in an intrinsically symmetrical
structure that increases in its intrinsic size by more than 25\%
at 7 mm.
\end{abstract}

\section{Introduction}

Sgr~A*, the extremely compact non-thermal radio source at the
galactic center, is a best candidate for a single super-massive
black hole (SMBH) from both the observational studies and the
theoretical models (cf. \cite{1}). The determination of the
orbital motions of those early-type stars within the vicinity of
Sgr~A* reveal a central dark mass about 4$\times10^6{\rm\
M_\odot}$ within a radius of about 45 AU \cite{2,3}. On the other
hand, the upper limit to the intrinsic proper motion of Sgr~A*
itself sets a lower limit of 0.4$\times10^6{\rm\ M_\odot}$ to the
mass directly associated with Sgr~A* \cite{4}. Now we can say that
we have known the mass of Sgr~A* to a great extent, the main
uncertainty being the exact distance to the galactic center. But
how big it is, and what it looks like, are different questions.

VLBI observations can provide the highest angular resolution
achievable with any astronomical instruments \cite{5}. At a
distance of 8 kc to Sgr~A* at the galactic center \cite{6}, an
angular separation of 1 mill-arcsecond (mas) corresponds to a
linear size of 8 AU. The Schwarzschild radius ($\rm R_{sc}$) of a
4$\times10^6{\rm\ M_\odot}$ SMBH is 1.2$\times$10$^{12}$ cm, or
0.08 AU or 10 micro-arcsecond ($\mu$as) in angular size. Indeed,
Sgr~A* is the closest SMBH candidate with the largest angular size
of its Schwarzschild radius on the sky, and thus one of the best
prime targets for the VLBI observational study.

\section{Millimeter-VLBI of Sgr~A*}

Attempts to measure the Sgr~A* structure with high-resolution VLBI
observations at centimeter (cm) wavelength have suffered from the
angular broadening caused by the diffractive scattering by the
ionized interstellar medium, which dominates the resultant images
with a $\lambda^2$-dependence apparent size, that is, $\Theta_{\rm
obs}=\Theta_{\rm obs}^{\rm 1 cm}\lambda^2$, where $\Theta_{\rm
obs}$ is the observed full-width at half-maximum (FWHM) in mas at
a wavelength $\lambda$ in cm, and equals $\Theta_{\rm obs}^{\rm 1
cm}$ at 1 cm. This tells us that the cm radio emission of Sgr~A*
comes from a region that extends to observers a much smaller angle
than the scattering angular size. However, following the
$\lambda^2$-dependent scattering law, the scattering angle
decreases very fast with the observing wavelength. So it can be
expected that VLBI observations at millimeter (mm) wavelength,
where the pure scattering angular size could become small enough
to be comparable to the intrinsic (finite) size of Sgr~A*, would
show deviations of the observed apparent size from the scattering
law.

Unlike the cm-VLBI observations, the mm-VLBI observations are
severely limited by the atmosphere due to the fact that Sgr~A* has
a southerly declination ($\sim- 30^\circ$) and most existing VLBI
antennas are located in the Northern Hemisphere. Therefore, most
of the observational data of Sgr~A* were taken at low elevation
angles ($10^\circ - 20^\circ$) where atmospheric effects are
substantial. Furthermore, the short and variable coherence time at
mm wavelength, combined with the compromised sensitivity due to
the high system temperature and the lower antenna efficiency of mm
VLBI antennas, seriously limits the high signal-to-noise-ratio
(SNR) detections of mm-VLBI observations of Sgr~A*. As a
consequence, there are large uncertainties in the results obtained
from the conventional imaging process, i.e. the self-calibration
technique for VLBI imaging, whose biggest drawback is
non-uniqueness when the SNR is poor.

\subsection{Revision of Scattering Law}
To circumvent the large uncertainty in the commonly used amplitude
calibration of VLBA observations of Sgr~A*, we have since 2001
developed a model fitting program in which the amplitude closure
relation is applied to minimize the calibration errors and thus to
improve the accuracy of the measurements \cite{7}. We then applied
this procedure to refine the apparent size measurements of the
first near-simultaneous VLBA+Y1 data \cite{8} observed at five
wavelengths (6, 3.6, 2, 1.35 cm, and 7 mm). By performing the
weighted least-squares fits to these sizes at different
wavelengths, we conclude that the best-fit two-dimensional
scattering structure is \cite{9}
\begin{equation}
\Theta_{\rm sca}^{\rm major} = (1.39 \pm 0.02)\lambda^2 ~,
\end{equation}
and
\begin{equation}
\Theta_{\rm sca}^{\rm minor} = (0.69\pm0.06)\lambda^2 ~,
\end{equation}
with a position angle of 80$^\circ$. Here, $\Theta_{\rm sca}^{\rm
major}$ and $\Theta_{\rm sca}^{\rm minor}$ are FWHM of major and
minor axes in mas. This newly revised wavelength-dependent
two-dimensional scattering structure is consistent with that from
directly fitting to the closure amplitudes of somewhat different
databases \cite{10}. Compared with previous one (cf. \cite{8}),
new scattering law shows an even smaller scattering angular size
along the East-West major axis direction.

\subsection{Detections of Intrinsic Size}
On November 3, 2002, we successfully carried out the first VLBA
imaging observation of Sgr~A* at its shortest wavelength of 3.5 mm
\cite{9}. Compared to the past 3.5 mm VLBI observations with the
CMVA \cite{11}, our VLBA observation, with its dynamic scheduling,
the highest possible recording rate and more frequent pointing
calibration, has for the first time determined its elliptical
structure in the radio emission at 3.5 mm, consistent with the
morphology seen at longer wavelengths. It should be pointed out
that dynamic observing is crucial to minimize the atmospheric
effects in the mm-VLBI observation of Sgr~A*. The exact observing
date is not fixed, but depends on the weather condition at most
antenna sites which usually are separated by a few thousand
kilometers. While observing, a frequent antenna pointing check
observations of compact and strong SiO maser sources, roughly in
every 10 minutes, is necessary to maintain the best reference
(offset) pointing and calibration.

The best fitted apparent structure from November 2002 VLBA
observations is 0.21$^{+0.02}_{-0.01}$ mas by
0.13$^{+0.05}_{-0.13}$ mas with a position angle 79$^{+12}_{-33}$
degrees. The major axis size is well determined while the fitted
size of the minor axis has a quite big error mainly due to the
poor resolution along the North-South direction. By subtracting in
quadrature the scattering angle from Eq. (1), we obtain, for the
first time, an intrinsic size of 0.126$\pm$0.017 mas along the
major axis at 3.5 mm. In September 2003, we performed the second
3.5 mm VLBA imaging observation of Sgr~A*, which confirms the
results in November 2002 (Shen et al. 2006 in preparation).

Similarly, two 7 mm VLBA+GBT (Green Bank Telescope) observations
with the highest recording rate were dynamically scheduled in
March 2004 with the resultant averaged sizes of the major and
minor axes of 0.724$\pm$0.001 mas and 0.384$\pm$0.013 mas,
respectively with position angle 80.6$^{+0.5}_{-0.6}$ degrees. The
inclusion of GBT can improve the resolution in North-South
direction by a factor of 3 \cite{7}, and actually this is the
first VLBI observations to make use of the GBT at 7 mm. The
results are consistent with the previous ones, but with much
better accuracies. A significant deviation of measured size from
the scattering angle along the major axis indicates an intrinsic
size of 0.268$\pm$0.010 mas for the major axis \cite{9},
consistent with the previously reported detection of intrinsic
size at 7 mm \cite{11}.

\subsection{Temporal Variation in the Structure of Sgr~A*}
On May 31, 1999, Sgr~A* was observed simultaneously at three 7 mm
bands (39, 43 and 45 GHz). It shows a larger deviation in the
apparent source size from scattering angle than that observed a
week ago on May 23, 1999 \cite{9}. Such a deviation is significant
at about 3$\sigma$ along the minor axis direction. Thus, for the
first time we can estimate an intrinsic size along the minor axis
of 0.359$\pm$0.095 mas. The derived intrinsic size for the major
axis is 0.334$\pm$0.042 mas. Therefore, within the error bars, the
intrinsic structure detected at 7 mm on May 31, 1999 could be
treated as a circular. Compared with the previously detected major
axis size of 0.268 mas, such an intrinsic structure has increased
in its size by at least 25\% at 7 mm.

\section{Discussion}
We have sampled a zone of the SMBH closer to the event horizon
than ever before, by detecting the intrinsic size of Sgr~A* to be
only 1 AU, or 12.6 R$_{\rm sc}$. Assuming an intrinsically
spherical structure of Sgr~A* and using the lower limit to the
mass of Sgr~A* of 4$\times10^5{\rm\ M_\odot}$ from the absolute
proper motion work \cite{4}, we can easily derive a lower limit to
the dark mass density of Sgr~A* of 6.5$\times$10$^{21}{\rm\
M_\odot}~{\rm pc}^{-3}$. This is the highest mass density ever
measured in any SMBH candidates, and thus strongly supports the
SMBH nature of Sgr~A*.

It is intriguing that the detected intrinsic size at 3.5 mm is
about twice the diameter of the shadow caused by the strong
gravitational bending of light rays \cite{12}. Regardless of the
exact emission model, the characteristic diameter of shadow is
always about 5R$_{\rm sc}$, which is nearly 50 $\mu$as for Sgr~A*.
So, Sgr~A* will be the most important target for the future sub-mm
VLBI experiment to test the general relativity in the strong field
regime. Such kind of shadow, if confirmed, would be the most
direct evidence for the existence of SMBH. The success of earlier
single baseline 1.3 mm VLBI experiments \cite{13} has already
demonstrated the feasibility of capturing an image of the shadow
around the edge of Sgr A* at sub-mm wavelengths in the near
future.

From the two well determined intrinsic size at 7 \cite{9, 11} and
3.5 mm \cite{9}, we can derive a $\lambda^\beta$-dependence of the
intrinsic size with $\beta=1.09^{+0.34}_{-0.32}$ using the
two-point fit. This intrinsic size versus wavelength relation is
consistent with the two lower limits of 0.02 and 0.008 mas at 1.3
and 0.8 mm, respectively, from the absence of refractive
scintillation \cite{14}.

Such a wavelength dependence of the intrinsic major axis size
provides a strong constraint on emission and accretion models for
Sgr A*. It explicitly rules out explanations other than those
models with stratified structure. It has been shown that the
radiatively inefficient accretion flow (RIAF) model, after taking
into account the interstellar scattering, can predict the observed
sizes at both 7 and 3.5 mm quite well \cite{15}. The extrapolated
size of emitting region at 1 mm will reach the last stable orbit
(LSO) radius of 3R$_{\rm sc}$ for a non-rotating SMBH. For a
rotating SMBH, the radius of LSO could be only 0.5R$_{\rm sc}$.
Therefore, a break in the wavelength-dependent intrinsic size is
inevitable with the decreasing wavelength, which can be used to
constrain the spin of SMBH.

The radio flux density of Sgr~A* is known to be variable at all
the observable wavelengths on time scales of days to months. The
variability appears to be more pronounced at shorter wavelengths
with a relatively large amplitude fluctuation. For example, an
intra-day variability (IDV) in the 2 mm flux density of Sgr~A* has
been found by the Nobeyama Millimeter Array (NMA) observations,
indicating a very compact emitting region \cite{16}. The
correlation between flux density at 7 mm and spectral index
suggests that the variation at the short radio wavelengths is
intrinsic rather than due to the interstellar scintillation
\cite{17}. As such, it seems inevitable that the variability in
the radio flux density of Sgr~A* would be accompanied by the
structural change. Our simulation seems to suggest that the
detected structural variability on May 31, 1999 could be related
to an outburst/flare occurred at about 50R$_{\rm sc}$ from the
central SMBH (Shen et al. in preparation).

\section{Conclusions}

We have revised the scattering law, which gives a smaller
scattering size along the East-West major axis direction. With the
dynamic scheduling, we successfully carried out two 3.5 mm VLBA
imaging observations, which reveal a consistent East-West
elongated apparent structure. We also performed two-epoch new 7 mm
VLBA+GBT observations with a better sensitivity and resolution.
These new data show that the intrinsic size has come to play with
the scattering effect at both 3.5 and 7 mm. The inferred intrinsic
size is about 1 and 2.1 AU at 3.5 and 7 mm, respectively. The
derived extraordinarily high (dark) mass density of Sgr~A*
strongly argues that Sgr~A* is indeed a super-massive black hole.

\ack This work was supported in part by the National Natural
Science Foundation of China (grant 10573029). Z.-Q. Shen
acknowledges support by the One-Hundred-Talent Program of the
Chinese Academy of Sciences. In addition to the author, people who
are involved in the project are P. T. P. Ho (ASIAA), M.-C. Liang
(Caltech), K. Y. Lo (NRAO), M. Miyoshi (NAOJ) and J.-H. Zhao
(CfA).

\end{document}